\documentclass[12pt]{article}
\usepackage{color,enumerate}
\usepackage{amsfonts,amssymb,amsmath}
 \usepackage{theorem,cite}
\usepackage{hyperref}

\newtheorem{definition}{Definition}[section]

\newtheorem{theorem}[definition]{Theorem}

\theorembodyfont{\rmfamily}
\newtheorem{rmk}{Remark}[section]

\numberwithin{equation}{section}
\definecolor{brique}{rgb}{.9,.2,0}
\definecolor{blvert}{rgb}{0,.8,.85}
\definecolor{vertcl}{rgb}{0,1,.7}

\newcommand{\bea}{\begin{eqnarray}}
\newcommand{\eea}{\end{eqnarray}}
\newcommand{\beano}{\begin{eqnarray*}}
\newcommand{\eeano}{\end{eqnarray*}}
\newcommand{\beq}{\begin{equation}}
\newcommand{\eeq}{\end{equation}}
\newcommand{\nonu}{\nonumber \\}

\newcommand{\hs}[1]{\hspace{#1 mm}}

\newcommand{\eps}{\epsilon}

    \def\cE{{\cal E}}    
    \def\cH{{\cal H}}

\def\cV{{\cal V}}    \def\cW{{\cal W}}

\newcommand{\wt}[1]{\widetilde{#1}}
\newcommand{\mb}[1]{\hs{4}\mbox{#1}\hs{4}}

\newcommand{\half}{\frac{1}{2}}

\newcommand{\ket}[1]{| #1 \rangle}

\DeclareMathOperator{\res}{Res}
\begin{document}
\renewcommand{\thefootnote}{\fnsymbol{footnote}}
\newpage
\pagestyle{empty}
\setcounter{page}{0}
%%%%%%%%%%%%%%%%%%%%%%%%%%%%%%%%
%%%%%  HEADINGS POUR DRAFT  %%%%%%

% \markright{\today\dotfill DRAFT\dotfill }
% \pagestyle{myheadings}

% \section{titre}
% \hspace{-1cm}\lapth

% \vspace{20mm}

\null
\vfill

\begin{center}

\vfill

 {\LARGE  {\sffamily 
Matrix Coordinate Bethe Ansatz: \\[1.2ex]
Applications to XXZ and ASEP models }}\\[1cm]

\vfill
  
{\large N. Crampe$^{ab,}$\footnote{nicolas.crampe@univ-montp2.fr}, 
E. Ragoucy$^{c,}$\footnote{ragoucy@lapp.in2p3.fr}
and D. Simon$^{d,}$\footnote{damien.simon@upmc.fr}}\\
\vfill 
{\large $^a$ Universit\'e Montpellier 2, Laboratoire Charles Coulomb UMR 5221,
F-34095 Montpellier, France \\[.242cm]
$^b$ CNRS, Laboratoire Charles Coulomb, UMR 5221, F-34095 Montpellier
}\\[.6cm]
{\large $^{c}$ Laboratoire de Physique Th{\'e}orique LAPTH\\[.242cm]
 CNRS and Universit{\'e} de Savoie.\\[.242cm]
   9 chemin de Bellevue, BP 110, F-74941  Annecy-le-Vieux Cedex, 
France. }\\[.6cm]
{\large $^{d}$ LPMA, Universit\'e Pierre et Marie Curie,\\[.242cm]
Case Courrier 188, 4 place Jussieu,
75252 Paris Cedex 05, France}
\end{center}
\vfill\vfill

\begin{abstract}
We present the construction of the \emph{full} set of eigenvectors of the 
open ASEP and XXZ models
with special constraints on the boundaries. The method combines both recent 
constructions 
of coordinate Bethe Ansatz and the old method of  matrix Ansatz specific to the ASEP. 
This "matrix coordinate Bethe Ansatz" can be viewed as a non-commutative 
coordinate Bethe 
Ansatz, the non-commutative part being related to the algebra appearing in 
the matrix Ansatz.
\end{abstract}

\vfill
\rightline{\texttt{hal-00603322}}
\rightline{\texttt{arXiv:1106.3264}}
\rightline{LAPTH-018/2011}

\newpage
\pagestyle{plain}
\renewcommand{\thefootnote}{\arabic{footnote}}
\setcounter{footnote}{0}
\section{Introduction}

Questions of integrability of models in statistical and quantum mechanics have been 
much more studied for periodic systems than for open systems, for which the numbers of 
particles and excitations may vary. However, open boundary conditions have become 
central in non-equilibrium physics, for which exactly solvable models are needed to 
explore new features. Fast advances have been made in the eighties to generalize 
Yang-Baxter equations to the open case (see \cite{sklyanin}). 
However, although the boundaries which preserve the integrability have 
been classified quite easily \cite{dvgr2,inko},
the computation of the eigenvalues and the eigenvectors for the non-diagonal 
boundaries is a tricky problem. Indeed,
 no standard method exists to 
diagonalize completely Sklyanin-type transfer matrices with 
non-diagonal boundary matrices. Recently, progresses have been made in this direction 
\cite{Cao, galleas, nepo, tq, BK, simon09, xxz-bound}, but the problem 
is far from been solved. 
The present paper addresses this 
question for a simple model that is used both in statistical mechanics and in spin chains.

The Asymmetric Simple Exclusion Process (ASEP) is the simplest model of transport of 
hard-core particles along a one-dimensional lattice, that exhibits non-trivial behaviours. 
Each site of the lattice may be occupied by one particle or empty. On the boundaries, 
two reservoirs add or remove particles with different rates and create a non-zero flux 
particle from one to the other. For a lattice of $L$ sites, the state space is 
$2^L$-dimensional 
and it can be mapped exactly to a system of $L$ spin $1/2$, the role of the Markov 
transition 
matrix being held by the XXZ spin chain Hamiltonian. The role of the reservoirs corresponds 
in this case 
to non-diagonal magnetic fields on the boundaries. Although all the coefficients have very 
different interpretations and may be complex or real depending on the model, 
the mathematical 
structure remains the same and integrability techniques can be applied in both cases. 
In the present paper, ASEP notations are chosen, without any loss of generality if 
one assumes 
that the coefficients may become complex. The detailed model is described in section
\ref{sec:definitions}, 
as well as previous integrability results on this model.

Although the models are identical, there exists at least one method that has been developed 
and applied only to the ASEP case: the Matrix Ansatz (or DEHP method) \cite{dehp}. 
The reason 
for this fact is that it requires the \emph{a priori} knowledge of one eigenvalue to 
build the 
corresponding eigenvector; it is precisely the case for stochastic models for which 
the existence 
of invariant measures is known. Attempts to establish parallels between Bethe and 
Matrix Ans\"atze produced 
various partial results \cite{alcaraz,golinellimallick} (see however 
\cite{hosho} for the case of periodic boundary conditions) but never a full understanding 
of the exact 
relation between both approaches. The present paper shows how to build coordinate Bethe 
Ansatz 
eigenvectors from a Matrix Ansatz-type vacuum state in section \ref{sec:matrixansatz}. 

The plan of the paper is the following. We start in section \ref{sec:definitions} 
with a brief summary 
on open ASEP models. Then, in section \ref{sec:matrixansatz}, we present our method, 
that can be viewed 
as a non-commutative coordinate Bethe Ansatz. The non-commutative part is based on 
the algebra 
appearing in the matrix Ansatz. We conclude in section \ref{sec:conclu}. 
Appendices \ref{sec:BC} 
and \ref{sec:resF} gather technical results needed in our construction. Finally, 
appendix \ref{sec:rep} is devoted to the study of finite dimensional representations of the 
matrix Ansatz algebra.

\section{ASEP Hamiltonians and constraints on boundaries\label{sec:definitions}}

The Markov transition matrix for the open ASEP model is given by
\beq \label{eq:hamasep}
W=\widehat K_1+K_L+\sum_{j=1}^{L-1}w_{j,j+1}
\eeq
where the indices indicate the spaces on which the following matrices 
act non trivially
\beq
w=\left(
\begin{array}{c c c c}
 0 & 0 & 0 & 0\\
 0 &-q & p & 0\\
 0 & q &-p & 0\\
 0 & 0 & 0 & 0
\end{array}
\right)
\mb{,}
\widehat K=\left(
\begin{array}{c c}
 -\alpha & \gamma e^{-s} \\
 \alpha e^{s} & -\gamma 
\end{array}
\right)
\mb{and}
K=\left(
\begin{array}{c c}
 -\delta & \beta  \\
 \delta  & -\beta 
\end{array}
\right)
\label{def:w12}
\eeq

It is well-established \cite{ssa,esr,dGE2005,dGE} that this ASEP model is 
related by a similarity transformation to the so-called integrable open XXZ model
\cite{sklyanin}. Some care must be taken with the properties of the matrix $W$. 
In the case of the ASEP with 
$s=0$, the matrix is stochastic but not hermitian. In the context of the ASEP, $s$ 
is the parameter of the 
generating function of the current of particles and, for $s\neq 0$, the matrix $W$ 
is even not stochastic. 
In the general case encompassing both ASEP and XXZ, there is no stochastic nor 
self-adjoint property for $W$ 
and left and right eigenvectors are generically different.

Among the whole set of integrable boundaries, 
there is one very special subset of constraints for which
 the resolution is simplified. This subset, in the notation of the ASEP model, 
is determined 
by the following constraints between the boundary parameters $\alpha$, 
$\beta$, $\gamma$, $\delta$ and $s$ but also 
of the bulk parameters $p$ and $q$ \cite{dGE2005,dGE,simon09, xxz-bound}
\begin{eqnarray}
\label{eq:constraint1}
&& c_\epsilon(\alpha,\gamma) c_{\epsilon'}(\beta,\delta) = e^s 
\left(\frac{p}{q}\right)^{L-1-n}\mb{for} \epsilon,\,\epsilon'=\pm
\mb{and} n\in\{0,1,2,\ldots,L-1\}\\
&&c_+(x,y)=y/x \mb{and} c_-(x,y)=-1\,.
\end{eqnarray}
The four possible choices for the signs $\epsilon,\epsilon'$ and the 
corresponding constraint are summarized in table \ref{tab:cons1}, 
as well as the corresponding notation used
for the Markov matrix $W$ when the constraints are satisfied.

\begin{table}[h b t]
\begin{centering}
\begin{tabular}{| c | c || c ||c|}
 \hline
 $c_\epsilon(\alpha,\gamma)$ 
&$c_{\epsilon'}(\beta,\delta)$
& Constraints & W
\\  
\hline
 $c_+(\alpha,\gamma)=\frac{\gamma}{\alpha} $  & $ 
c_+(\beta,\delta)=\frac{\delta}{\beta}$ & 
$ 
\frac{\alpha\beta}{\gamma\delta}e^s\left(\frac{p}{q}\right)^{L-1-n}=1$&
$W_n^{++}$\\  
\hline
 $c_-(\alpha,\gamma)=-1 $  &$ c_-(\beta,\delta)=-1$        
& $ e^s\left(\frac{p}{q}\right)^{L-1-n} = 1 $  &
$W_n^{--}$               \\  
\hline
 $ c_-(\alpha,\gamma)=-1$ &  $ 
c_+(\beta,\delta)=\frac{\delta}{\beta}$    
& $ -\frac{\beta}{\delta}e^s\left(\frac{p}{q}\right)^{L-1-n} = 
1$      &
$W_n^{-+}$                  \\  
\hline
  $ c_+(\alpha,\gamma)=\frac{\gamma}{\alpha}$ & $  
c_-(\beta,\delta)=-1$
 &  $ - \frac{\alpha}{\gamma}e^s\left(\frac{p}{q}\right)^{L-1-n} = 
1$     &
$W_n^{+-}$                  \\
\hline
\end{tabular}
\caption{Possible values for the parameters and the 
constraints imposed by \eqref{eq:constraint1} where $n$ takes integer values between 
$0$ and $L-1$.
\label{tab:cons1}}
\end{centering}
\end{table}
Similarly to $W$, we denote by $K^{\epsilon,\epsilon'}_n$ and 
$\widehat{K}^{\epsilon,\epsilon'}_n$ the boundary
matrices $K$ and $\widehat K$ when the corresponding constraint (\ref{eq:constraint1}) 
is satisfied.
In the ASEP, the parameters $\alpha$, $\beta$, $\gamma$ and $\delta$
are positive. 
Then, only the first two constraints in the table \ref{tab:cons1} may be considered as 
relevant. 
However, we treat also the last two sets since they become relevant when we map the ASEP 
problem to the XXZ model. 

By numerical investigations in \cite{neporav}, it has been established that the whole 
spectrum 
of $W_n^{\epsilon,\epsilon'}$ is given by two different types of Bethe equations.
In previous papers \cite{simon09, xxz-bound}, we succeeded in computing the eigenvalues 
and the left or right 
eigenvectors corresponding to only one type of Bethe equation via a generalization 
of the coordinate Bethe Ansatz \cite{bethe}.  It seems 
that this generalized coordinate Bethe Ansatz is not enough to obtain the missing cases. 
An advantage of this method consists in giving an interpretation of the number $n$ 
entering in the constraint:
it is the maximal number of pseudo-excitations in the Ansatz.

For $W_{L-1}^{--}$ (i.e. $e^s=1$), a second method, now called matrix Ansatz 
(or DEHP method),
 has been developed in \cite{dehp} to find the 
second part of the spectrum. We give the outlines of the historical method in 
subsection \ref{sec:dehp}.
Then, we present the new results in subsection \ref{sec:man} which consists in a 
generalization of the matrix Ansatz. It allows us to
obtain the second part of the spectrum for any 
$W_n^{\epsilon,\epsilon'}$, i.e. for any subset of constraints.

In \cite{simon09, xxz-bound}, we proved also that for 
a \emph{new} type of constraints the generalized coordinate Bethe 
Ansatz also provides one part of the spectrum 
\begin{eqnarray}
\label{eq:constraint2}
&&c^*_\epsilon(\alpha,\gamma)c^*_{\epsilon'}(\beta,\delta) = 
e^s\left(\frac{p}{q}\right)^{n}\mb{for} \epsilon,\,\epsilon'=\pm
\mb{and} n\in\{0,1,\ldots,L-1\}\\
\label{eq:defcs}
&&c^*_\pm(u,v) = \frac{p-q+v-u \pm 
\sqrt{(p-q+v-u)^2+4uv}}{2u} \label{eq:defcstar}
\,.
\end{eqnarray}
The four possible choices for the signs $\epsilon,\epsilon'$ and the 
corresponding constraints are summarized in table \ref{tab:cons2}. We denote by 
$\cW^{\epsilon,\epsilon'}_n$
 the matrix $W$ when one 
constraint (\ref{eq:constraint2}) is satisfied.
\begin{table}[h b t]
\begin{centering}
\begin{tabular}{| c | c || c ||c|}
 \hline
 $\epsilon$ &$\epsilon'$
& Constraints & $W$
\\  
\hline
 $+$  & 
$+$ & 
$ c^*_+(\alpha,\gamma)c^*_+(\beta,\delta)
=e^s\left(\frac{p}{q}\right)^{n}$& $\cW^{++}_n$
\\  
\hline
 $-$  & 
$-$ & 
$ c^*_-(\alpha,\gamma)c^*_-(\beta,\delta)
=e^s\left(\frac{p}{q}\right)^{n}$& $\cW^{--}_n$\\  
\hline
 $+$  & 
$-$ & 
$c^*_+(\alpha,\gamma)c^*_-(\beta,\delta)
=e^s\left(\frac{p}{q}\right)^{n}$& $\cW^{+-}_n$\\  
\hline
 $-$  & 
$+$ & 
$c^*_-(\alpha,\gamma)c^*_+(\beta,\delta)
=e^s\left(\frac{p}{q}\right)^{n}$& $\cW^{-+}_n$\\  
\hline
\end{tabular}
\caption{Possible values for the parameters and the 
constraints imposed by \eqref{eq:constraint2}, where $n$ is an integer between $0$ and $L-1$.
\label{tab:cons2}}
\end{centering}
\end{table}

As for the matrix $W^{\epsilon,\epsilon'}_n$, in \cite{xxz-bound}, we computed
 one part of the spectrum for $\cW^{\epsilon,\epsilon'}_n$ Markov matrices thanks to the 
 generalized coordinate Bethe Ansatz. We believe that a very similar procedure as 
the one presented 
in following section \ref{sec:man} should hold in this case when we transform the matrix 
as in \cite{xxz-bound}. We will not give in detail the computations in the present paper.

\section{Matrix coordinate Bethe Ansatz\label{sec:matrixansatz}}

\subsection{Matrix Ansatz without excitation\label{sec:dehp}}

In this subsection, we present the method introduced in \cite{dehp} to get 
the eigenvector with vanishing eigenvalue
(i.e. the steady state or the invariant measure) for the Markov process with $s=0$.
This state corresponds exactly to the part of eigenspace we do not obtain from the 
coordinate Bethe Ansatz
for $W_{L-1}^{--}$ in \cite{simon09,xxz-bound}.

A particular care is required since it involves various operators acting on 
\emph{different} vector spaces. 
The physical state space, written all along the paper $\cH$, is $2^L$-dimensional and the 
canonical basis can be indexed by the occupation numbers
$\tau_i\in\{0,1\}$ (resp. spin $s_i\in\{-\half,\,\half\}$ in the XXZ language) 
on each site $i$. All the vectors of $\cH$ are written in the ket notation $|\cdot\rangle$ 
and all the dual 
vectors are written in the bra form $\langle \cdot |$.

The matrix Ansatz states that the steady state $|\Phi\rangle$ of 
the ASEP with $s=0$ has components given by:
\begin{equation}
\label{eq:oldmatrixansatz}
\langle \tau_1\tau_2\ldots\tau_L | \Phi \rangle = \langle\langle V_1 
| \prod_{1\leq i \leq L}^{\longrightarrow}\left( \tau_i D+ 
(1-\tau_i)E \right) | V_2 \rangle\rangle\,,
\end{equation}
where the arrow means that the product has to be build from left to 
right when the index $i$ increases. One has for example $\langle 
011001|\Phi\rangle= \langle\langle V_1 | EDDEED | V_2 
\rangle\rangle$ for $L=6$. The non-commuting matrices $D$ and $E$ act on an 
abstract auxiliary vector space $\mathcal{V}$, that is different from $\cH$. 
The vector $| V_1 
\rangle\rangle$ lies in this space $\mathcal{V}$ (as any vector written with a ket 
notation $|\cdot\rangle\rangle$), 
whereas the vector 
$\langle\langle V_2 |$ is in its dual $\cV^*$ (as any vector written with 
a bra notation $\langle\langle \cdot |$). 
Eq.\eqref{eq:oldmatrixansatz} expresses the components of $|\Phi\rangle\in\cH$ as a 
scalar product between vectors on 
the abstract auxiliary space $\cV$.

One checks that $|\Phi\rangle$ is an eigenvector of 
$W_{L-1}^{--}$ with vanishing eigenvalue if the two matrices $D$ et $E$ acting on 
$\cV$ and the two 
boundary vectors satisfy the commutation rules:
\begin{subequations}
\label{matrixansatz}
\begin{align}
qED-pDE &= D+E \label{eq:comED}\,,
\\
\langle\langle V_1 | (\alpha E - \gamma D+1) &= 0 \label{eq:mag}\,,
\\
(\beta D -\delta E+1) | V_2 \rangle\rangle &=0 \label{eq:mad}\,.
\end{align}
\end{subequations}
These three relations reduce quadratic relations in $D$ and $E$ to linear expressions 
in $D$ and $E$ in the 
bulk and linear relations in $D$ and $E$ to scalar expressions on the boundaries; it 
allows one to determine 
recursively all the components of the eigenvector $|\Phi\rangle$ and it does not need 
an explicit 
representation of the algebra. An irrelevant minus sign relatively to the standard 
notation for the matrix 
Ansatz has been introduced in \eqref{matrixansatz} for later convenience. In particular, 
the matrix Ansatz gives 
an easy access to correlation functions with standard 
transfer matrix techniques \cite{dehp}.

To prove easily that $|\Phi\rangle$ is an eigenvector of $W_{L-1}^{--}$, we firstly 
rewrite $|\Phi\rangle$ as follows
\begin{equation}\label{eq:Phi0}
 |\Phi\rangle=\langle\langle V_1 | 
\left(\begin{array}{c}
 E\\
D
\end{array}\right)
\otimes
\left(\begin{array}{c}
 E\\
D
\end{array}\right)\otimes\dots\otimes
\left(\begin{array}{c}
 E\\
D
\end{array}\right)
| V_2 \rangle\rangle
\end{equation}
The convention used in this notation will be used all through the paper and corresponds 
to the following interpretation: 
a vector $\begin{pmatrix} E \\ D \end{pmatrix}$ has operator entries instead of scalar 
ones (two dimensional 
module over the endomorphisms of $\cV$), the tensor product of such elements is an element 
of a $2^L$-dimensional 
module over the endomorphisms of $\cV$, whose components are products of size $L$ of 
operators $E$ and $D$ with 
the usual tensor rule:
\begin{equation}
 \begin{pmatrix}
   a \\ b
 \end{pmatrix}
\otimes 
\begin{pmatrix}
 a' \\ b'
\end{pmatrix}
\equiv \begin{pmatrix}
        aa' \\ ab' \\ ba' \\ bb'
       \end{pmatrix}
\;.
\end{equation}
Let us emphasize on the order of the non-commuting operators in the vector of the r.h.s. 
of the previous equation. 
The right action of $|V_2 \rangle\rangle\in\cV$ produces a $2^L$-dimensional vector whose 
components are 
vectors of $\cV$, that is further projected through the left action of 
$\langle\langle V_1 |\in \cV^*$ to 
a $2^L$-dimensional vector whose components are complex numbers and is thus identified 
to $\cH$. For example,
the following vector
\begin{equation}
\langle\langle V_1 |  \begin{pmatrix}
   a \\ b
 \end{pmatrix}
\otimes 
\begin{pmatrix}
 a' \\ b'
\end{pmatrix}| V_2 \rangle\rangle
\equiv \begin{pmatrix}
 \langle\langle V_1 |  aa'| V_2 \rangle\rangle \\ \langle\langle V_1 |ab'| V_2 \rangle\rangle 
\\ 
\langle\langle V_1 |ba'| V_2 \rangle\rangle \\ \langle\langle V_1 | bb'| V_2 \rangle\rangle
 \end{pmatrix}
\;
\end{equation}
is a 4-component vector with complex entries.

Secondly, using definitions (\ref{def:w12}) and equations (\ref{matrixansatz}), we show that
\begin{eqnarray}
&& \omega\left(\begin{array}{c}
 E\\  D
\end{array}\right)
\otimes
\left(\begin{array}{c}
 E\\  D
\end{array}\right)=
\left(\begin{array}{c}
 E\\  D
\end{array}\right)
\otimes
\left(\begin{array}{c}
 1\\  -1
\end{array}\right)-
\left(\begin{array}{c}
 1\\  -1
\end{array}\right)
\otimes
\left(\begin{array}{c}
 E\\  D
\end{array}\right)\label{eq:bu01}\\
&&\widehat{K}_{L-1}^{--}\langle\langle V_1 | \left(\begin{array}{c}
 E\\  D
\end{array}\right)=\langle\langle V_1 |\left(\begin{array}{c}
 1\\  -1
\end{array}\right)\qquad\text{and}\quad
{K}_{L-1}^{--}\left(\begin{array}{c}
 E\\  D
\end{array}\right)|V_2\rangle\rangle=-\left(\begin{array}{c}
 1\\  -1
\end{array}\right)|V_2\rangle\rangle\qquad\label{eq:KV}
\end{eqnarray}
The first equation encodes four relations on the endomorphisms $D$ and $E$ and each 
equation on the second 
line two relations between vectors of $\cV$ or $\cV^*$. Since this type of result is 
central in this paper
and to be pedagogical,
we explain in detail the computation for the second relation in (\ref{eq:KV})
\begin{eqnarray}
 {K}_{L-1}^{--}\left(\begin{array}{c}
 E\\  D
\end{array}\right)|V_2\rangle\rangle&=&
% {K}_{L-1}^{--}
\left(\begin{array}{cc}
-\delta & \beta \\ \delta & -\beta 
\end{array}\right)
\left(\begin{array}{c}
 E|V_2\rangle\rangle\\  D|V_2\rangle\rangle
\end{array}\right)
=
\left(\begin{array}{c}
(\beta D -\delta E)|V_2\rangle\rangle\\ (\delta E -\beta  D)|V_2\rangle\rangle
\end{array}\right)
\label{eq:exemple-lg1}\\
&=&-\left(\begin{array}{c}
 |V_2\rangle\rangle\\  -|V_2\rangle\rangle
\end{array}\right)=
-\left(\begin{array}{c}
 1\\  -1
\end{array}\right)|V_2\rangle\rangle
\label{eq:exemple-lg2}
\end{eqnarray}
where to go from (\ref{eq:exemple-lg1}) to (\ref{eq:exemple-lg2}) we 
used (\ref{eq:mad}).
Let us remark that we denote by $1$ the identity 
operator acting on $\cV$.

Finally, remarking that the matrices $\omega$, $K$ and 
$\widehat K$ do not act in the auxiliary space $\cV$ 
where are $\langle\langle V_1 |$ and $|V_2\rangle\rangle$, we obtain telescopic terms that
can be simplified to get
\begin{equation}
 W_{L-1}^{--}|\Phi\rangle=0
\end{equation}

The goal of this paper consists in generalizing (\ref{eq:Phi0}) to obtain also eigenvalues 
and eigenvectors for all
the  $W_{n}^{\epsilon,\epsilon'}$. For that purpose, we add some pseudo-excitations above 
the previous 
eigenvector as it is done in usual Bethe Ansatz.
Before giving the matrix Ansatz with $n$ pseudo-excitations, we need to introduce some 
vectors playing the role
 of these excitations as well as some of their properties. 

\subsection{Pseudo-excitations and properties} 

Let us introduce the notation:
\begin{equation}
 [u]=\frac{1-u}{q-p}
\end{equation}
and the vectors
\begin{equation}
\boldsymbol{|\omega(u)\rangle}=\left(
\begin{array}{c}
 E-[u]\\
uD+[u]
\end{array}
\right) 
\qquad
\boldsymbol{|t(u)\rangle}=\left(
\begin{array}{c}
u\\
-1
\end{array}
\right)
\qquad
\boldsymbol{|\overline{t}(u)\rangle}=(p-q)\left(
\begin{array}{c}
0\\
uD+[u]
\end{array}
\right)
\end{equation}
where the notation $\boldsymbol{|.\rangle}$ (ket in bold) stands for 2-component vectors 
in which
the entries are operators on $\cV$ (to differentiate them from vectors ${|.\rangle}\in\mathcal{H}$ introduced in section \ref{sec:dehp}, which contain complex numbers and are obtained after projection with $\langle\langle V_1 |$ and $|V_2\rangle\rangle$).

If the non-commuting matrices $D$ and $E$ satisfy relation (\ref{eq:comED}), we get the 
following relations
\begin{subequations}
\label{eq:bu}
 \begin{align}
w \boldsymbol{|\omega(u)\rangle}\otimes \boldsymbol{|\omega(u)\rangle}=&
\boldsymbol{|\omega(u)\rangle}\otimes \boldsymbol{|t(u)\rangle}
-\boldsymbol{|t(u)\rangle}\otimes \boldsymbol{|\omega(u)\rangle}
\label{eq:bu1}\\
w \boldsymbol{|\omega(v)\rangle}\otimes \boldsymbol{|\omega(v\frac{q}{p})\rangle}=&
\boldsymbol{|\omega(v)\rangle}\otimes \boldsymbol{|\overline{t}(v\frac{q}{p})\rangle}
-\boldsymbol{|\overline{t}(v)\rangle}\otimes \boldsymbol{|\omega(v\frac{q}{p})\rangle}\label{eq:bu2}\\
w \boldsymbol{|\omega(u)\rangle}\otimes \boldsymbol{|\omega(v)\rangle}=&
-p\boldsymbol{|\omega(u)\rangle}\otimes \boldsymbol{|\omega(v)\rangle}
+p\boldsymbol{|\omega(v)\rangle}\otimes \boldsymbol{|\omega(u\frac{q}{p})\rangle}\nonumber\\
&-\boldsymbol{|t(u)\rangle}\otimes \boldsymbol{|\omega(v)\rangle}+\boldsymbol{|\omega(u)\rangle}\otimes\boldsymbol{|\overline{t}(v)\rangle}\label{eq:bu3} \\
w \boldsymbol{|\omega(v)\rangle}\otimes \boldsymbol{|\omega(u\frac{q}{p})\rangle}=&
-q\boldsymbol{|\omega(v)\rangle}\otimes \boldsymbol{|\omega(u\frac{q}{p})\rangle}+q\boldsymbol{|\omega(u)\rangle}\otimes \boldsymbol{|\omega(v)\rangle}\nonumber\\
&+\boldsymbol{|\omega(v)\rangle}\otimes \boldsymbol{|t(u\frac{q}{p})\rangle}
-\boldsymbol{|\overline{t}(v)\rangle}\otimes \boldsymbol{|\omega(u\frac{q}{p})\rangle}
\label{eq:bu4}
 \end{align}
\end{subequations}
where $u$ and $v$ are still arbitrary numbers that will be fixed later. The proof of 
relations (\ref{eq:bu}) is straightforward: 
one projects each relation 
on the four components then one uses the definition of the $[u]$ numbers as well as the 
commuting relation (\ref{eq:comED}).
Let us remark that relation (\ref{eq:bu1}) for $u=1$ is similar to relation (\ref{eq:bu01}).
Relations (\ref{eq:bu}) deal with the bulk part of $W$. We study now these vectors on the 
boundaries.

Let us introduce the following functions
\begin{equation}
 \lambda_+(u,v)=0 \qquad\text{and}\quad \lambda_-(u,v)=-u-v\;.
\end{equation}
In the following section, we impose the following relation on the right boundary, for 
$\eps=\pm$,
\begin{equation}\label{eq:bd}
 K \boldsymbol{|\omega(u)\rangle}|V_2\rangle\rangle =
\lambda_\eps(\beta,\delta)\boldsymbol{|\omega(u)\rangle}|V_2\rangle\rangle
-\boldsymbol{|t(u)\rangle}|V_2\rangle\rangle
\end{equation}
where, as explained previously, 
$\boldsymbol{|\omega(u)\rangle}|V_2\rangle\rangle$ means that $|V_2\rangle\rangle$ is 
right-applied 
to each entry of $\boldsymbol{|\omega(u)\rangle}$. 
This relation provides generally two different constraints ($K$ is a two-by-two matrix) 
on vectors of $\cV$ except
 if $u=-1/c_{-\eps}(\beta,\delta)$. In that case, the remaining constraint becomes 
\begin{equation}
(\beta D-\delta E +1)|V_2\rangle\rangle=0
\quad\text{for}\quad
 u=-1/c_{-\eps}(\beta,\delta)\,.
\end{equation}
We recover the relation (\ref{eq:mad}) for both values of 
 $\lambda_\eps(\beta,\delta)$.    

Similarly, on the left boundary, we impose the following relation
\begin{equation}\label{eq:bg}
 \widehat{K} \langle\langle V_1|\boldsymbol{|\omega(u)\rangle} =
\lambda_\eps(\alpha,\gamma)\langle\langle V_1|\boldsymbol{|\omega(u)\rangle}
+\langle\langle V_1| \boldsymbol{|t(u)\rangle}
\end{equation}
where the notation $\langle\langle V_1|\boldsymbol{|\omega(u)\rangle}$ means that 
$\langle\langle V_1|$ is left-applied 
to each operator 
entry of $\boldsymbol{|\omega(u)\rangle}$. 
With a special choice of $u$, the two constraints reduce again to only 
one, given by
\begin{equation}
\langle\langle V_1|\Big(\alpha e^s(E-[u])-\frac{\gamma}{u e^s}(uD+[u])+1\Big)=0
\quad\text{for}\quad
 u=-e^{-s} c_{-\eps}(\alpha,\gamma)\,.
\end{equation}

\subsection{Matrix Ansatz with excitations\label{sec:man}}

Let us define the tensor product (for modules defined on the endomorphisms of $\cV$)  
of vectors 
$\boldsymbol{|\omega(u)\rangle}$ over the 
sites $i$ to $j$ and write it as
\begin{equation}
\label{eq:def:Omega}
\boldsymbol{\ket{\Omega(u)}}_i^j= 
\boldsymbol{\ket{\omega(u)}}_i\boldsymbol{\ket{\omega(u)}}_{i+1}\ldots
\boldsymbol{\ket{\omega(u)}}_j
\,.
\end{equation}
We now fix an integer $n$ and introduce the state with $n-m$ 
excitations at the ordered positions $1\leq x_{m+1}< \ldots < x_n 
\leq L$ defined as the $(\mathbb{C}^2)^{\otimes L}$-vector in $\cH$ with the projections:
\bea
&&\ket{x_{m+1},\ldots,x_n} \ =\ \left(\sqrt{\frac{q}{p}}\right)^{(x_{m+1}-1)+\ldots+(x_n-1)} 
\label{eq:def:state}\\
&&\times
\langle\langle V_1|~\boldsymbol{\ket{\Omega(u_{m+1})}}_1^{x_{m+1}-1}
\boldsymbol{\ket{\omega(v_{m+1})}}_{x_{m+1}}
\boldsymbol{\ket{\Omega(u_{m+2})}}_{x_{m+1}+1}^{x_{m+2}-1} 
.. \boldsymbol{\ket{\omega(v_n)}}_{x_n}
\boldsymbol{\ket{\Omega(u_{n+1})}}_{x_n+1}^L~ |V_2\rangle\rangle\,.
\nonumber
\eea
The overall factor $\sqrt{q/p}$ is introduced only in order 
to normalize the Bethe roots. The coefficients $u_m$ and $v_m$ are 
related through the recursions:
\begin{equation}
\label{eq:recursion:ui:vi}
 u_{m+1} = \frac{q}{p} u_m, \quad v_{m+1} =  \frac{q}{p} v_m\,,
\end{equation}
and the initial coefficients $u_1$ and $v_1$ are still arbitrary. 
These vectors correspond to states where $m$ excitations have left 
the 
system (through the left boundary). To clarify the notation, the state
with no excitation corresponds to $m=n$ and is given by
\begin{equation}
\ket{\emptyset}=\langle\langle V_1|~ \boldsymbol{\ket{\Omega(u_{n+1})}}_{1}^{L}~ 
|V_2\rangle\rangle\,,
\end{equation}
and, for $u_{n+1}=1$, we recover the state $|\Phi\rangle$ defined in (\ref{eq:Phi0}). 

We need also to introduce some other definitions concerning the set on which we are going 
to sum in our Ansatz.
The set $G_m$ is a full set of representatives of the coset 
$BC_n/BC_m$ ($G_0=BC_n$, by convention) and $BC_{m}$ is the $B_{m}$ Weyl group, generated by  
transpositions $\sigma_{j}$, $j=1,\ldots,m-1$ and the reflection $r_{1}$  
(for details, see appendix \ref{sec:BC}).
It acts on a vector $\boldsymbol{k}=(k_1,\dots,k_n)$ of 
$\mathbb{C}^n$:
\begin{equation}
 \boldsymbol{k}_{r_1}=(-k_1,k_2,\dots,k_n)\quad,\quad
 \boldsymbol{k}_{\sigma_j}=(k_1,\dots,k_{j+1},k_j,\dots,k_n)\,.
\end{equation}
We introduce also the following truncated vector $\boldsymbol{k}^{(m)}_{g}$, 
for $0\leq m\leq n$ and $g\in G_m$,
\begin{equation}
\boldsymbol{k}^{(m)}_{g} =(k_{g(m+1)},\dots,k_{gn})\;.
\end{equation}

We are now in position to state the main result of this paper which provides eigenvalues 
and eigenstates
of $W_{L-1-n}^{\epsilon,\epsilon'}$ i.e. the matrix $W$ 
with constraint $
 c_\epsilon(\alpha,\gamma) c_{\epsilon'}(\beta,\delta) = e^s 
\left(\frac{p}{q}\right)^{n}$.

\begin{theorem}\label{th1}
The vector
\begin{equation}
\label{eq:ansatz}
 \ket{\Phi_n}=\sum_{m=0}^n\ \sum_{x_{m+1}<\dots<x_n}\ \sum_{g\in G_m}\
A_g^{(m)}\ e^{i\boldsymbol{k}^{(m)}_g.\boldsymbol{x}^{(m)}}\ 
|x_{m+1},\dots,x_n\rangle\,,
\end{equation}
is an eigenstate of $W_{L-1-n}^{\epsilon,\epsilon'}$ 
with eigenvalue
\begin{equation}\label{eq:E}
 \cE_{L-1-n}^{\epsilon,\epsilon'}=\lambda_{-\epsilon}(\alpha,\gamma)
+\lambda_{-\epsilon'}(\beta,\delta)+
\sum_{j=1}^n\Lambda(e^{ik_j})\mb{where} 
\Lambda(z)=\sqrt{pq}\left(z+\frac{1}{z}\right)-p-q
\end{equation}
if the following relations are fulfilled:
\begin{enumerate}[a)]
 \item\label{eq:1} The coefficients $u_1$ and $v_1$ entering in definition 
(\ref{eq:def:state}) are
\begin{equation}\label{eq:uv}
 u_1=-e^{-s} c_{\epsilon}(\alpha,\gamma)\mb{and}
v_1=\frac{1}{2\delta}\left(\frac{p}{q}\right)^{m-1}
\Big(p-q+\delta-\beta\pm\sqrt{(\beta-\delta+q-p)^2+4\beta\delta}\Big)
\end{equation}
\item\label{eq:2} The non-commuting elements $E$ and $D$ obey
\begin{eqnarray}
 &&qED-pDE=E+D\\
&&(\beta D-\delta E +1)|V_2\rangle\rangle=0\\
&&\langle\langle V_1|\Big(\alpha e^s(E-[u_1])-\gamma e^{-s}(D-[1/u_1])+1\Big)=0\label{eq:VED}\,.
\end{eqnarray}
 \item\label{eq:3} The coefficients $A_g^{(m)}$ verify
\begin{eqnarray}
\label{eq:S}
&&A^{(0)}_{g\sigma_j}
=S\left(e^{ik_{gj}},e^{ik_{g(j+1)}}\right)~A^{(0)}_{g}\,,\\
\label{eq:recuT}
&& A_g^{(m)}=T^{(m)}(e^{ik_{g1}},\dots,e^{ik_{gm}})A_g^{(m-1)}\,,
\end{eqnarray}
where 
 \begin{equation}
 S(z_{1},z_{2})=-\frac{a(z_{1},z_{2})}{a(z_{2},z_{1})}
\mb{with}
a(z_{1},z_{2})=\frac{i}{z_{1}z_{2}-1}\left(\left(\sqrt{\frac{q}{p}}
+\sqrt{\frac{p}{q}}\right)z_{2}-z_{1}z_{2}-1\right)\,,
\label{def:alphax}
\end{equation}
and
\begin{eqnarray}
&&T^{(m)}(z_{1},\dots,z_{m}) =\frac{D_1^{(m-1)}}{p_1(z_{m}) 
V_1^{\epsilon}(z_{m})}
\frac{z_m^2-1}{\prod_{j=1}^{m-1}a(z_{m},z_j)a(z_{j},1/z_m)} \,,
\\
&&\label{eq:Dm}
D_1^{(m-1)}=
\frac{v_m}{v_m-u_{m+1}}\Big(\alpha e^s v_m +\gamma-\alpha+p-q-\frac{\gamma}{e^s\ v_m}\Big)
\,,\\
&&V_1^{\pm}(z) =\Lambda(z)
+ (\lambda_{\mp}(\alpha,\gamma)+\gamma) \Big(1-\frac{1}{z}\sqrt{\frac{p}{q}}\Big)
+ (\lambda_{\mp}(\alpha,\gamma)+\alpha)\Big(1-\frac{1}{z}\sqrt{\frac{q}{p}}\Big)\,,
\quad\label{def:V1x} \\
&&p_1(z) =z+  \frac{1}{\sqrt{pq}} \frac{pu_2-q v_1}{v_1-u_2} \,.
\label{def:coeff:r}
\end{eqnarray}
 \item\label{eq:4} The pseudo-momentum $k_j$ must satisfy the following Bethe equations, 
for $1\leq j \leq n$,
\begin{equation}
\label{eq:bethe}
 \prod_{\substack{\ell=1 \\ \ell\neq j}}^n 
S(e^{ik_\ell},e^{ik_j})S(e^{-ik_j},e^{ik_\ell})
=e^{2iLk_j}
\frac{V_1^{\epsilon}(e^{ik_{j}})V_L^{\epsilon'}(e^{ik_{j}})}
{V_1^{\epsilon}(e^{-ik_{j}})V_L^{\epsilon'}(e^{-ik_{j}})}\,,
\end{equation}
with
\begin{equation}
V_{L}^{\pm}(z) = \Lambda(z)+
\left(\lambda_{\mp}(\beta,\delta)+\beta\right)
\Big(1-\frac 1z\sqrt{\frac qp}\Big)
+(\lambda_{\mp}(\beta,\delta)+\delta)\Big(1-\frac 1z\sqrt{\frac 
pq}\Big)\,.
\label{def:VLx}
\end{equation}
\end{enumerate}
\end{theorem}
Before giving the proof of this theorem in subsection \ref{sec:pr}, we make some remarks 
on the theorem:
\begin{rmk} \label{rm1}
The first relation in 
(\ref{eq:uv}) is equivalent, via constraint (\ref{eq:constraint1}), to 
\begin{equation}
 u_{n+1}=-1/c_{\epsilon'}(\beta,\delta)\;.
\end{equation}
The sign in the definition of $v_1$ is irrelevant. 
\end{rmk}

\begin{rmk}
The algebra generated by $E$ and $D$ is very closed to the one introduced in \cite{dehp}: 
the only difference lies in equation (\ref{eq:VED}).
We study the finite dimensional representations of this algebra in appendix \ref{sec:rep}, 
which gives intriguing relations with the second set of constraints \eqref{eq:constraint2}.
\end{rmk}

\begin{rmk}
A consequence of (\ref{eq:recuT}) (for $m=1$) and 
$A_{gr_1}^{(1)}=A_g^{(1)}$, is 
\begin{equation}
 A_{gr_1}^{(0)}
=\frac{T^{(1)}(e^{ik_{g1}})}{T^{(1)}(e^{-ik_{g1}})}A_g^{(0)}\,.
\end{equation}
This relation with (\ref{eq:S}) allow us to express $A_g^{(0)}$ for 
any $g\in BC_n$ in terms of $A_{1}^{(0)}$ 
(where the subscript $1$ stands for the unit of $BC_n$ group).
Finally, using recursively (\ref{eq:recuT}), we can express all the 
coefficients $A_g^{(m)}$ in terms of only $A_1^{(0)}$. This last 
coefficient is 
usually chosen such that the eigenfunction $|\Phi_n\rangle$ be normed.
\end{rmk}

\begin{rmk}\label{rm4}
In our previous work \cite{xxz-bound}, we found via the coordinate Bethe Ansatz the
 eigenfunctions of $W_n^{\epsilon,\epsilon'}$
with eigenvalues
\begin{eqnarray}\label{eq:schCBA}
&& \wt\cE_n^{\epsilon,\epsilon'}
=\lambda_{\epsilon}(\alpha,\gamma)+\lambda_{\epsilon'}(\beta,\delta)+
\sum_{j=1}^n\Lambda(e^{ip_j})
\end{eqnarray}
Remark the change of signs in the index of both $\lambda$ in these eigenvalues
(\ref{eq:schCBA}) 
in comparison with (\ref{eq:E}). 
The pseudo-momentum $p_j$ satisfy the following Bethe equations
\begin{equation}
\label{eq:betheCBA}
 \prod_{\substack{\ell=1 \\ \ell\neq j}}^n 
S(e^{ip_\ell},e^{ip_j})S(e^{-ip_j},e^{ip_\ell})
=e^{2iLp_j}
\frac{V_1^{-\epsilon}(e^{ip_{j}})V_L^{-\epsilon'}(e^{ip_{j}})}
{V_1^{-\epsilon}(e^{-ip_{j}})V_L^{-\epsilon'}(e^{-ip_{j}})}\,,\qquad
j=1,2,\ldots,n
\end{equation}
By numerical investigations \cite{neporav}, it has been conjectured that 
the set of eigenvalues given by (\ref{eq:schCBA}), (\ref{eq:betheCBA}) with
 $n$ pseudo-excitations and by (\ref{eq:E}), (\ref{eq:bethe}) with $L-1-n$ 
pseudo-excitations
give the complete spectrum of $W_n^{\epsilon,\epsilon'}$. Therefore, our 
previous results \cite{xxz-bound} with
the results of this paper seem to provide the complete spectrum for 
$W_n^{\epsilon,\epsilon'}$ as well as 
the associated
eigenstates.
\end{rmk}

\subsection{Proof of the main theorem \ref{th1}\label{sec:pr}}

To prove the theorem, we show that the following equation holds
\begin{equation}\label{eq:schp}
W_{L-1-n}^{\epsilon,\epsilon'} |\Phi_n\rangle = \cE_{L-1-n}^{\epsilon,\epsilon'} 
|\Phi_n\rangle\;.
\end{equation}
The proof is very similar to the one we performed in \cite{xxz-bound} for the 
generalized coordinate Bethe Ansatz, except that entries of 
$\boldsymbol{|\omega(\cdot)\rangle}$ vectors are now non-commuting operators. 
It consists in projecting equation (\ref{eq:schp}) on the different $\ket{x_{m+1},\ldots,x_n}$
and to prove that each projection is true if the conditions \ref{eq:1})-\ref{eq:4}) 
of the theorem hold.
We write only the projections 
leading to independent relations (one can check that the remaining 
ones 
do not lead to new relations).

\paragraph{On $\boldsymbol{|x_{1},\dots,x_n\rangle}$ for 
$\boldsymbol{(x_{1},\dots,x_n)}$ generic} (i.e. $1<x_1$, $x_n<L$ and 
$1+x_j<x_{j+1}$)\\

Before performing this projection, let us remark that, 
using relations (\ref{eq:bu}), (\ref{eq:bd}) and (\ref{eq:bg}) as well as the conditions 
\ref{eq:2}) of the theorem, 
we can show that 
\begin{eqnarray}
W_{L-1-n}^{\epsilon,\epsilon'} |x_{1},\dots,x_n\rangle&=&
\Big(\lambda_{-\epsilon}(\alpha,\gamma)+\lambda_{-\epsilon'}(\beta,\delta)\Big)
\ket{x_{1},\ldots,x_n}\\
&&+\sum_{j=1}^n\Big[(-p-q)\ket{x_{1},\ldots,x_n}+\sqrt{pq}(\ket{\ldots,x_j+1,\ldots}
+\ket{\ldots,x_j-1,\ldots})\Big]
\nonumber
\end{eqnarray}
We remind that relations (\ref{eq:bd}) and (\ref{eq:bg}) are valid only if
\begin{equation}
 u_1=-e^{-s} c_{\epsilon}(\alpha,\gamma)\mb{and}u_{n+1}=-\frac{1}{c_{\epsilon'}(\beta,\delta)}
\end{equation}
which is the first relation in \ref{eq:1}) (see also remark \ref{rm1}).
In addition of that, the recursion relation between the $u$'s (\ref{eq:recursion:ui:vi}) 
implies the constraint
$c_\epsilon(\alpha,\gamma) c_{\epsilon'}(\beta,\delta) = e^s 
\left(\frac{p}{q}\right)^{n}$ which is the constraint for $W_{L-1-n}^{\epsilon,\epsilon'}$.

Finally, the projection on $|x_{1},\dots,x_n\rangle$ (for generic
$x_{1},\dots,x_n$) of equation (\ref{eq:schp}) holds if the energy takes the form (\ref{eq:E}).

\paragraph{On $\boldsymbol{|x_{1},\dots,x_n\rangle}$ with 
$\boldsymbol{x_{j+1}=1+x_{j}}$} (and 
$x_1,\dots,x_{j-1},x_{j+2},\dots,x_n$ generic)\\

Using (\ref{eq:bu2}) and this projection, we get a relation (\ref{eq:S}) between 
$A^{(0)}_{g}$ and $A^{(0)}_{g\sigma_j}$.
As expected, the expression of the scattering matrix is similar to the periodic case since the 
boundaries 
are not involved in this process.

\paragraph{On $\boldsymbol{|x_{m+1}\dots,x_n\rangle}$} 
($x_{m+1},\dots,x_n$ generic and $m\geq 1$)\\

Before performing this projection, we need to know how the left boundary matrix 
$\widehat K$ acts on 
the vectors $|\omega(v_m)\rangle$ and $|\omega(u_m)\rangle$. By direct computation, 
using relation (\ref{eq:VED}),
we show that
\begin{eqnarray}
&&\hspace{-1.5cm} \widehat K_{L-1-n}^{\epsilon,\epsilon'} \langle\langle V_1| 
\boldsymbol{|\omega(v_m)\rangle}=
\widetilde{\Lambda}_1^{(m-1)}\langle\langle V_1|\boldsymbol{|\omega(v_m)\rangle}+
D_1^{(m-1)}\langle\langle V_1|\boldsymbol{|\omega(u_{m+1})\rangle}
+\langle\langle V_1|\boldsymbol{|\overline{t}(v_{m})\rangle}
\\
&&\hspace{-1.5cm}\widehat K_{L-1-n}^{\epsilon,\epsilon'} 
\langle\langle V_1|\boldsymbol{|\omega(u_m)\rangle}=
{\Lambda}_1^{(m-1)}\langle\langle V_1|\boldsymbol{|\omega(u_m)\rangle}+
C_1^{(m-1)}\langle\langle V_1|\boldsymbol{|\omega(v_{m-1})\rangle}
+\langle\langle V_1|\boldsymbol{|{t}(v_{m})\rangle}
\end{eqnarray}
where the same notational convention is used as for \eqref{eq:bg} and
where $D_1^{(m-1)}$ is defined in (\ref{eq:Dm}) and
\begin{eqnarray}
 \widetilde{\Lambda}_1^{(m-1)}&=&
\frac{1}{u_{m+1}-v_m}\Big(\alpha e^s v_m u_{m+1}+(\gamma+p-q)v_m-\alpha u_{m+1}
-\gamma e^{-s}\Big)
\\
{\Lambda}_1^{(m-1)}&=&\frac{(1-e^s u_m)(\alpha v_{m-1}+\gamma e^{-s})}{u_m-v_{m-1}}
\\
C_1^{(m-1)}&=&\frac{(1-e^s u_m)(\alpha u_{m}+\gamma e^{-s})}{v_{m-1}-u_m}
\end{eqnarray}
Using these relations, we get finally that the projection on 
${|x_{m+1}\dots,x_n\rangle}$ provides the following constraints, for any $g\in G_m$,
\begin{equation}\label{eq:1g}
 D_1^{(m-1)}\sum_{h\in H_m}A_{gh}^{(m-1)}e^{ik_{gh(m)}}
+(\Lambda_1^{(m)}-\lambda_{-\epsilon}(\alpha,\gamma)-
\sum_{j=1}^{m}\Lambda(e^{ik_{gj}}))A_g^{(m)}=0
\,.
\end{equation}
where $H_m=BC_m/BC_{m-1}$ (see Appendix \ref{sec:BC}). 

We are going to demonstrate that this last constraint (\ref{eq:1g}) holds if relations 
(\ref{eq:S}) and
(\ref{eq:recuT}) are true\footnote{This demonstration is similar to the one done in 
\cite{xxz-bound} but 
we give here again for completeness of the present proof.}. We start by remarking that 
a consequence 
of latter equations (\ref{eq:S}) and (\ref{eq:recuT}) is
\begin{equation}
\label{eq:Sm}
 A_{g\sigma_\alpha}^{(m)}=
A_{g}^{(m)}\times\begin{cases}
 1 &1\leq \alpha \leq m-1\,,\\
\frac{T^{(m)}(e^{ik_{g1}},\dots,e^{ik_{gm-1}},e^{ik_{gm+1}})}
{T^{(m)}(e^{ik_{g1}},\dots,e^{ik_{gm}})}
S(e^{ik_{gm}},e^{ik_{gm+1}})& \alpha= m\,,\\
S(e^{ik_{g\alpha}},e^{ik_{g\alpha+1}})& \alpha\geq m+1\,.
\end{cases}
\end{equation}
Then, using again (\ref{eq:recuT}) to express now
$A_g^{(m)}$ in terms of $A_g^{(m-1)}$
and using (\ref{eq:Sm}) to express $A_{gh}^{(m-1)}$ ($h\in H_m$) in terms of 
$A_g^{(m-1)}$, relation 
(\ref{eq:1g}) becomes
the functional relation
\begin{eqnarray}\label{eq:ss1}
&&\sum_{j=1}^{m}\left[\frac{z_jV^{\epsilon}_1(z_j)p_1(z_j)}{z_j^2-1}
\prod_{\substack{\ell=1\\ \ell\neq j}}^{m}
a(z_j,z_\ell)a(z_\ell,\frac{1}{z_j})
+\frac{z_j}{1-z_j^2}V^{\epsilon}_1(\frac{1}{z_j})p_1(\frac{1}{z_j})
\prod_{\substack{\ell=1\\ \ell\neq 
j}}^{m}a(\frac{1}{z_j},z_\ell)a(z_\ell,z_j)
\right]
\nonu
&&\qquad =\lambda_{-\epsilon}(\alpha,\gamma)-\Lambda_1^{(m)}+\sum_{j=1}^m\Lambda(z_j)\,,
\end{eqnarray}
where $z_j$ stands for $\exp(ik_{gj})$ and the functions are defined 
in (\ref{def:alphax})-(\ref{def:V1x}).
To prove this last relation (\ref{eq:ss1}), let us introduce the 
following function
\begin{equation}\label{eq:F}
 F^{(m)}(z)=\sqrt{pq}\frac{V^{\epsilon}_1(z)p_1(z)}
{\Lambda(z)\left(2z-\sqrt{\frac{p}{q}}-\sqrt{\frac{q}{p}}\right)}
\prod_{\ell=1}^ {m} a(z,z_\ell)a(z_\ell,\frac{1}{z})\,.
\end{equation}
Then, one can prove that (\ref{eq:ss1}) is equivalent to $\sum_{\text{residues of $F^{(m)}$}} 
F^{(m)}(z)=0$ 
(see appendix \ref{sec:resF} for the complete list of its residues),
which finishes to demonstrate that constraint (\ref{eq:1g}) is verified 
if relations (\ref{eq:S}) and
(\ref{eq:recuT}) are true.

\paragraph{On $\boldsymbol{|1,x_{m+1}\dots,x_n\rangle}$} 
($x_{m+1},\dots,x_n$ generic and $m\geq1$)\\

This projection provides a second relation between the coefficients 
from the level $m-1$ and $m$.
We obtain the following constraint, for any $g\in G_m$,
\begin{eqnarray}\label{eq:2g}
\sum_{h\in H_m}\Big(\sqrt{pq}e^{ik_{ghm}}+\widetilde{\Lambda}_1^{(m-1)}
-\lambda_{-\epsilon}(\alpha,\gamma)-q
-\sum_{j=1}^{m}\Lambda(e^{ik_{gj}})\Big)
 A_{gh}^{(m-1)}e^{ik_{ghm}}
+C_1^{(m)}A_g^{(m)}=0\,.
\end{eqnarray}
We are going to prove that this constraint is satisfied if relations (\ref{eq:S}) and
(\ref{eq:recuT}) following similar demonstration as previously.
Using the previous projection (\ref{eq:1g}) (already proven) then 
relations (\ref{eq:S}) and (\ref{eq:recuT}), 
we prove that projection (\ref{eq:2g}) becomes the following 
functional relation 
\begin{eqnarray}\label{eq:ss2}
&&\sqrt{pq}\sum_{j=1}^{m}\left[\frac{z_j^2}{z_j^2-1} V_1^{\epsilon}(z_j)p_1(z_j)
\prod_{\substack{\ell=1\\ \ell\neq j}}^{m}a(z_j,z_\ell)a(z_\ell,\frac{1}{z_j})
+\frac{V^{\epsilon}_1(\frac{1}{z_j})p_1(\frac{1}{z_j})}{1-z_j^2}
\prod_{\substack{\ell=1\\ \ell\neq j}}^{m}a(\frac{1}{z_j},z_\ell)a(z_\ell,z_j)
\right]\nonumber\\
&&=\Big(\sum_{j=1}^{m}\Lambda(z_j)
+\lambda_{-\epsilon}(\alpha,\gamma)-\widetilde\Lambda_1^{(m-1)}+q\Big)
\Big(\sum_{j=1}^{m}\Lambda(z_j)+\lambda_{-\epsilon}(\alpha,\gamma)-\Lambda_1^{(m)}\Big)
-C_1^{(m)}D_1^{(m-1)}\,.\qquad\
\end{eqnarray}
The function to consider is now 
$G^{(m)}(z)=\sqrt{pq}\,z\,F^{(m)}(z)$.
Finally, we prove that functional relation (\ref{eq:ss2}) is equivalent to 
$\sum_{\text{residues of $G^{(m)}$}} G^{(m)}(z)=0$ 
(see appendix \ref{sec:resF} for the computation of the residues). 

\paragraph{On $\boldsymbol{|x_{1}\dots,x_{n-1},L\rangle}$} 
($x_{1}\dots,x_{n-1}$ generic)\\

To perform this projection, we need to know the action of the right boundary on one 
pseudo-excitation
\begin{equation}
K_{L-1-n}^{\epsilon,\epsilon'} \boldsymbol{|\omega(v_n)\rangle} |V_2\rangle\rangle=
\widetilde{\Lambda}_L \boldsymbol{|\omega(v_n)\rangle} |V_2\rangle\rangle
-\boldsymbol{|\overline{t}(v_n)\rangle} |V_2\rangle\rangle
\end{equation}
where the same convention is used as for \eqref{eq:bd} and 
where
$ \widetilde{\Lambda}_L=\delta(v_n-1)$
and $v_n=\left(\frac{q}{p}\right)^{n-1}v_1$ with $v_1$ given by (\ref{eq:uv}).

Then we can prove that this last independent projection holds if the so-called Bethe 
equations (\ref{eq:bethe})
are satisfied.

\section{Conclusion\label{sec:conclu}}

The previous sections present the construction of the \emph{full} set of eigenvectors of 
the ASEP and the XXZ spin chain 
with special constraints on the boundaries. The method combines both recent constructions 
of coordinate Bethe Ansatz for 
the same set of constraints \cite{simon09,xxz-bound} and the old method of the matrix 
Ansatz \cite{dehp} specific to the ASEP. 
Although computations have been showed only one set of special constraints, the construction 
should be transposed without effort 
to the second set of special constraints discovered in \cite{xxz-bound}. Left eigenvectors 
are also very simple to build using 
the same tricks.

A first intriguing feature of the matrix coordinate Bethe Ansatz for the first set of 
special constraints 
(\ref{eq:constraint1}) is presented in appendix \ref{sec:rep}: finite dimensional 
representations of the matrices $D$ and $E$
 can be found only if the \emph{second} set of constraints \eqref{eq:constraint2} 
is satisfied. Up to now, however, we do not 
have simple explanations of this fact.

The matrix Ansatz has proven to be useful, at least in the case of zero excitations, 
for the study of correlation functions 
\cite{derrida-enaud} since it is reduced to standard transfer matrix techniques of 
one-dimensional statistical mechanics. 
The same question in the context of Bethe Ansatz is notably difficult and has lead to 
many different approaches such as the 
quantum inverse scattering method. It would be interesting to investigate whether the 
present formulation may simplify the study 
of correlation functions, either in the present case with boundaries or in the standard 
periodic case. 

The proofs presented here seem to indicate that the matrix Ansatz state plays the role 
of a new vacuum state, although it is 
highly non-trivial and do not factorize, in the context of the open XXZ spin chain. 
The standard coordinate Bethe Ansatz approach 
has been used but it would worth knowing if algebraic Bethe Ansatz could be adapted 
to obtain the same eigenvectors, as in \cite{Cao}.

Numerical evidence tends to show that the set of eigenvectors is now complete and 
gives a synthetic view of the $BC_n$ 
structure of the eigenstates. The constructions for the two sectors of the spectrum 
are similar on their structure but very 
different from the point of view of the reference vacuum state, although a physical 
interpretation for the ASEP gives some hints 
\cite{belitskyschuetz,schuetz-et-al}. One may hope that a further understanding of 
the passage from one sector to the other may 
allow one to couple both sectors and study ASEP and XXZ spin chains with boundaries 
out of the special constraints that allowed 
the present framework.

Finally, the matrix Ansatz approach was found to be useful for various stochastic models 
of particles with different types or impurities 
\cite{impurity,multispecies}, as well as for the study of tableaux in combinatorics 
\cite{combinatorics}, even if no Bethe Ansatz 
approach exists yet for these models. It would be interesting to know whether our approach 
may help for these other models, so that 
integrable system methods can be extended to them.

\appendix
\section{Weyl group $BC_n$ and cosets\label{sec:BC}}

The Weyl group $BC_n$ is generated by the set 
$\{r_1,\sigma_1,\dots,\sigma_{n-1}\}$ with the following 
constraints:\begin{equation}
 \sigma_j^2=1=r_1^2\quad,\quad \sigma_1 r_1\sigma_1 
r_1=r_1\sigma_1r_1\sigma_1\quad,\quad
\sigma_j\sigma_{j+1}\sigma_j=\sigma_{j+1}\sigma_j\sigma_{j+1}\,.
\end{equation}
The subgroup generated by $\{\sigma_1,\ldots,\sigma_{n-1}\}$ is just 
the symmetric group. We now consider its subgroups generated by 
$\{r_1,\sigma_1,\dots,\sigma_{m-1}\}$, $m\leq n$, 
which we identify with
$BC_m$.

For $g\in BC_n$, we then define the class $gBC_m=\{gh;\,h\in BC_m\}$, 
called a left coset. 
It is known that the set of all classes $gBC_n$, which is called 
$BC_n/BC_m$, forms a partition 
 $BC_n$: we can thus define $G_m$ as a full set of representative of 
$BC_n/BC_m$, such that one 
has the unique decomposition $BC_n=\bigcup_{g\in G_m} gBC_m$. We 
set, by convention, 
$G_0\sim BC_n$ and $G_n=\{1\}$.

The action of an element $g$ of $G_m$ on a vector
$\boldsymbol{k}^{(m)}=(k_{m+1},\dots,k_n)$ of $\mathbb{R}^{n-m}$
is given by $\boldsymbol{k}^{(m)}_g=(k_{g(m+1)},\dots,k_{g(n)})$. One 
checks that this action does not depend on the choice of the 
representative $g$, such that the action of $BC_n/BC_m$ is 
well-defined on $\mathbb{R}^{n-m}$ without further specifications.
This definition is useful because the set 
$\{\boldsymbol{k}^{(m)}_g|g\in G_m\}$ contains 
one and only one time the vector 
$(\eps_{i_1}k_{i_1},\dots,\eps_{i_{n-m}}k_{i_{n-m}})$
for any choice $\eps_j=\pm$, $1\leq i_j\leq n$ and $i_j\neq i_k$.
For example, $\{\boldsymbol{k}^{(n-1)}_g|g\in 
G_{n-1}\}=\{(k_n),(-k_n),(k_{n-1}),(-k_{n-1}),\dots,
(k_1),(-k_1)\}$.

Finally, we introduce $H_m$ which is a full set of representatives 
of the coset $BC_m/BC_{m-1}$ which may chosen as follows
\begin{eqnarray}
&& \{id,\sigma_{m-1},\sigma_{m-2}\sigma_{m-1},\dots,\sigma_1\dots\sigma_{m-2}\sigma_{m-1},
\\
&&\hspace{1cm}r_1\sigma_1\dots\sigma_{m-2}\sigma_{m-1},\sigma_1r_1\sigma_1\dots
\sigma_{m-2}\sigma_{m-1},\dots,
\sigma_{m-1}\dots\sigma_1r_1\sigma_1\dots\sigma_{m-2}\sigma_{m-1}\}\nonumber
\end{eqnarray}

\section{List of the residues of the function $F^{(m)}$ and $G^{(m)}$\label{sec:resF}}

We list in this appendix the residues of the function $F^{(m)}$ defined by (\ref{eq:F}) 
and $G^{(m)}$.
The residues of  $F^{(m)}$ are
\begin{eqnarray}
\res(F^{(m)}(z))\Big|_{z=z_j} &=& \frac{z_jV^{\epsilon}_1(z_j)p_1(z_j)}{z_j^2-1}
\prod_{\substack{\ell=1\\ \ell\neq j}}^{m}a(z_j,z_\ell)a(z_\ell,\frac{1}{z_j})\,,
\\
%%%%%%%%%%%%%%%%%%%
\res(F^{(m)}(z))\Big|_{z=1/z_j} &=& 
\frac{1}{z_j((1/z_j)^2-1)}V^{\epsilon}_1(\frac{1}{z_j})p_1(\frac{1}{z_j})
\prod_{\substack{\ell=1\\ \ell\neq j}}^{m}a(\frac{1}{z_j},z_\ell)a(z_\ell,z_j)\,,
\\
%%%%%%%%%%%%%%%%%%
\res(F^{(m)}(z))\Big|_{z=\sqrt{\frac{p}{q}}}
&=& \sqrt{pq}
\left(\frac{q}{p}\right)^{m} \frac{\lambda_{-\epsilon}(\alpha,\gamma)
+\alpha}{p-q}p_1(\sqrt{\frac{p}{q}})\,,
\\
%%%%%%%%%%%%%%%%%%%
\res(F^{(m)}(z))\Big|_{z=\sqrt{\frac{q}{p}}}
&=& \sqrt{pq}\left(\frac{p}{q}\right)^{m} 
\frac{\lambda_{-\epsilon}(\alpha,\gamma)+\gamma}{q-p}p_1(\sqrt{\frac{q}{p}})\,,
\\
%%%%%%%%%%%%%%%%%%%%
\res(F^{(m)}(z))
\Big|_{z=\frac{p+q}{2\sqrt{{p}{q}}}}
&=&\frac{\sqrt{pq}}{2(q-p)} \left( p-q+2\alpha-2\gamma\right) 
p_1(\frac{p+q}{2\sqrt{{p}{q}}})\,,
\\
%%%%%%%%%%%%%%%%%%%%%%%%
\res(F^{(m)}(z))\Big|_{z=\infty}
&=& 
-\sum_{\ell=1}^{m}\Lambda(z_\ell)
-\frac{\sqrt{pq}}{2} p_1(\frac{p+q}{2\sqrt{{p}{q}}})-\frac{1}{2}
(2\lambda_{-\epsilon}(\alpha,\gamma)+\alpha+\gamma)\,
\qquad
\end{eqnarray}
The 
residues of $G^{(m)}(z)$ at the point $z=z_{0}$ with $z_{0}\neq 0,\infty$ are easy
to compute: 
\beq
\res(G^{(m)}(z))\Big|_{z=z_{0}}=\sqrt{pq}\,z_{0}\, 
\res(F^{(m)}(z))\Big|_{z=z_{0}}\,. 
\eeq
Since 0 is not a pole, it remains 
to compute the residue at infinity:
\begin{eqnarray}
&& \res(G^{(m)}(z))\Big|_{z=\infty}
= \sum_{j,k=1}^{m} \Lambda(z_{j})\,\Lambda(z_{k}) + 2 
\left(\res(F^{(m)}(z))\Big|_{z=\infty}-\frac{p+q}{4} 
\right)\sum_{j=1}^{m}\Lambda(z_{j}) \nonu
&& - 
\frac{\sqrt{pq}}{2}p_1(\frac{p+q}{\sqrt{{p}{q}}})
\left(2\lambda_{-\epsilon}(\alpha,\gamma)+\alpha+\gamma+\frac{p+q}{2}\right)
-\frac{(p-q)(\alpha-\gamma)}{4}+ 
\frac{(p+q)^2}{8}\,.\quad
\end{eqnarray}

\pagebreak[4]
\section{Representations of the algebra generated by $E$ and $D$\label{sec:rep}}

In this appendix, we study the finite-dimensional irreducible
representations of the algebra we used to construct 
the matrix Ansatz since 
such representations emerge only for special constraints already encountered in the study 
of the open ASEP or XXZ spin chains. 
In the case without excitation, these types of representation have been studied previously in 
\cite{esr,mallicksandow}. We will follow similar proofs for the cases with excitations. 

To be self-contained 
in this appendix, we remind that the algebra needed in these cases is defined by
\begin{subequations}
\label{matrixansatz-app}
\begin{align}
&qED-pDE = D+E \label{eq:comED-app}\,,
\\
&(\beta D -\delta E+1) | V_2 \rangle\rangle =0 \label{eq:mad-app}\
\\
&\langle\langle V_1|\Big(\alpha e^s(E-[u])-\frac{\gamma}{u e^s}(uD+[u])+1\Big)=0
 \,,\label{eq:mag-app}
\end{align}
\end{subequations}
where $u=-e^{-s} c_{-\eps}(\alpha,\gamma)
=-(p/q)^n/{c_{-\epsilon'}(\beta,\delta)}$, 
 $\epsilon,\epsilon'\in\{\pm\}$, $c_+(x,y)=y/x$ and $c_-(x,y)=-1$.

In \cite{mallicksandow}, it is proved that, to satisfy the 'bulk' part (\ref{eq:comED-app}), 
the non-commuting elements
$E$ and $D$ must take the following form (up to similarity transforms)
\begin{eqnarray}
 D&=&\frac{1}{q-p}\sum_{j=1}^N \left(1+a\left(\frac{q}{p}\right)^{j-1}\right)E_{jj}\\
E&=&\frac{1}{q-p}\sum_{j=1}^N \left(1+\frac{1}{a}\left(\frac{p}{q}\right)^{j-1}\right)E_{jj}
+\frac{1}{q-p}\sum_{j=1}^{N-1}E_{j+1,j}
\end{eqnarray}
where $N$ is the dimension of the representation, $a$ is  a free parameter 
and $E_{ij}$ is the elementary matrix
with $1$ in the entry $(i,j)$ and $0$ otherwise.

Following the arguments of \cite{mallicksandow}, 
the 'boundary' conditions (\ref{eq:mad-app}) and (\ref{eq:mag-app}) imply that there exist
integers $k$ and $\ell$
 between $1$ and $N$ such that
\begin{eqnarray}
\label{eq:ca1}
&& \beta \left(1+a\left(\frac{q}{p}\right)^{k-1}\right) 
-\delta\left(1+\frac{1}{a}\left(\frac{p}{q}\right)^{k-1}\right) 
+q-p=0\\
\label{eq:ca2}
&&\alpha e^s\left(u+\frac{1}{a}\left(\frac{p}{q}\right)^{\ell-1}\right)
-\frac{\gamma}{u e^s}\left(1+au\left(\frac{q}{p}\right)^{\ell-1}\right)+q-p=0\;.
\end{eqnarray}
$N$ being the size of the irreducible representation, one gets the constraint $|k-\ell|=N-1$.
In the case $k=N>1$ and $\ell=1$, one has 
\begin{equation}
 \langle\langle V_1|=(1,0,\dots,0)\mb{and} |V_2\rangle\rangle =(0,\dots,0,1)^t
\quad\Rightarrow\quad \langle\langle V_1|D^j|V_2\rangle\rangle =0\,,\ 
\forall\ j\,.\label{eq:dj=0}
\end{equation}
Now, since (\ref{eq:comED-app}) implies that
\beq
q^n\,E\,D^n = (1+p\,D)^n\,E+\sum_{\ell=1}^{n}
q^{\ell-1}\,(1+p\,D)^{n-\ell}\,D^{\ell}\qquad \forall n\,,
\eeq
(\ref{eq:dj=0}) shows, together with (\ref{eq:mad-app}), that all words 
(in $E$ and $D$) vanish. 
Thus, this case must be excluded.

It remains the case $k=1$ and $\ell=N\geq1$ for which equations (\ref{eq:ca1}) and (\ref{eq:ca2}) become
\begin{eqnarray}
&& \beta a^2+(\beta-\delta+q-p)a-\delta=0\\
&&\alpha\left(\frac{e^{s}}{a}\left(\frac{p}{q}\right)^{N-1}\right)^2
+\left(\alpha u e^s-\frac{\gamma}{ue^s}+q-p\right)\left(\frac{e^{s}}{a}
\left(\frac{p}{q}\right)^{N-1}\right)
-\gamma=0\;.
\end{eqnarray}
Then, using definition (\ref{eq:defcs}), we get, for any choice for $u$,
\begin{equation}
 a=c^*_\tau(\beta,\delta)\mb{and}
\frac{e^{s}}{a}\left(\frac{p}{q}\right)^{N-1}=c^*_{\tau'}(\alpha,\gamma)
\mb{with} \tau,\,\tau'=\pm\,.
\end{equation}
Therefore, a finite dimensional representation exists if there exist two signs $\tau$ 
and $\tau'$
such that the following relation is true
\begin{equation}
c^*_\tau(\alpha,\gamma)\, c^*_{\tau'}(\beta,\delta)=e^s \left(\frac{p}{q}\right)^{N-1}
\end{equation}
where the functions $c^*_\pm$ have been defined in (\ref{eq:defcstar}).

For real parameters and $u,v>0$, one has 
$c^*_{-}(u,v)<0$ and $c^*_{+}(u,v)>0$ (since for $uv>0$ one has
$\sqrt{(p-q+v-u)^2+4uv}>|p-q+v-u|$). Thus, for ASEP models, only the 
constraints corresponding to $\tau=\tau'$ have to be considered (as 
it is the case for the first choice). We recognize in the framework of 
matrix Ansatz the same constraints (\ref{eq:constraint2}) that appear  
for coordinate Bethe Ansatz, although the relation between both approaches are very different.

% \section{biblio}

\end{document}